\documentclass[prl,aps,twocolumn,showpacs,floatfix]{revtex4-1}
\usepackage{epsfig}
\usepackage{amssymb}
\usepackage [fleqn]{amsmath}
\usepackage{amsfonts}
\usepackage{graphicx,color}

\newcommand {\LSCO} {${\rm La_{2-x}Sr_xCuO_4}$}
\newcommand {\tc} {$\rm{T_c}$}

\newcommand {\be} {\begin{equation}}
\newcommand {\ee} {\end{equation}}

\newcommand{\G} {\tilde\Gamma}

\begin{document}

\title{Zero Energy States at a Normal--Cuprate-Superconductor Interface Probed by Shot Noise}
\author{O.~Negri$^{a,b}$, M.~Zaberchik$^{a,b}$, A.~Keren$^{a}$, M.~Reznikov$^{a,b}$}
\affiliation{
$^{a}$ Department of Physics, $^{b}$ Solid State Institute,
Technion--Israel Institute of Technology, Haifa, 3200003, Israel\\
}

\begin{abstract}
We report measurements of the current noise generated by the optimally doped, x=0.15, Au-\LSCO{} junctions. For  high transmission junctions on (110) surface, we observed split zero-bias conductance peak (ZBCP), accompanied by enhanced shot noise. We attribute the enhanced noise to the Cooper pair transport through the junction. The ZBCP disappears and the noise decreases to the one expected for the charge $e$ with heating at temperatures well below \tc, and at voltages much smaller than the bulk superconducting gap, setting a new energy scale of $0.5$\,mV. We attribute this scale to the existence of an  $id_{xy}$ or $is$ order parameter at the sample surface.
\end{abstract}

\date{\today}

\maketitle

Since the seminal work of Andreev\,\cite{andreev}, it has been understood that the current through a normal metal--superconductor (NS) interface at voltages below \tc{} and zero temperature is carried by Cooper pairs. This current, being partitioned by the interface, should generate shot noise with effective charge $2e$\,\footnote{ Shot noise generated in a {\em normal diffusive metal} by Cooper pairs reflected from a transparent NS interface was observed in\,\cite{jehl99,reuletAndreev03}.}. Despite seemingly simple prediction, such doubled shot noise was observed only recently\,\cite{Heiblum12} in a single channel wire connected to a BCS superconductor. In a normal metal--BCS superconductor junction, the low-voltage probability of pair transmission is $\Gamma_p=\Gamma_e^2$, where $\Gamma_e$ is the single-electron transmission probability through the junction interface. At a finite temperature, the Cooper pair current competes with the single-particle one, proportional to $\Gamma_e$ and, for a sufficiently low-transmission junction, such that  $\Gamma_e\ll \exp(-\Delta/k_BT)$, $\Delta$ being the gap, single-electron transport dominates.

The situation changes dramatically for a superconductor with momentum-dependent gap $\Delta({\bf k})$. In the case $\Delta(-k_\perp)=-\Delta(k_\perp)$, where $k_\perp$ is the momentum component in the direction normal to the NS interface, surface states with the energy around the middle of the superconductive gap are formed\,\cite{bruder,Hu,kashiwaya}. This leads to a zero-bias conduction peak (ZBCP) with $\Gamma_e$-dependent width, observed in tunneling experiments on d-wave High-Temperature Superconductors (HTSC)\,\cite{sharoni02,sharoni03}. Treatment, similar to the one of the BTK paper\,\cite{btk} for BCS superconductors, leads to a prediction\,\cite{kashiwaya} of reflectionless pair transmission ($\Gamma_p=1$) at zero voltage.
Based on this expectation, it was predicted that the current in ZBCP region would generate no shot noise\,\cite{zhu1999shot,kashiwaya_noise}. Indeed, spectral density of the shot noise at zero temperature is given\,\cite{Khlus1987,Lesovik1989,Yurke1990} by: $S=2qI(1-\Gamma_p)$, where $q$ is the effective carrier charge.

The papers\,\cite{Hu,kashiwaya,zhu1999shot,kashiwaya_noise} treat an ideal case of a translationally-invariant NS interface, at which carrier reflection is specular. In practice, however, the reflection on an NS interface is diffusive. Moreover, the surface of a superconductor is far from being perfect: it contains oxygen vacancies as in the case of YBCO, or native oxygen layer, or just a normal layer of HTSC with substantial impurity scattering. Such nonidealities destroy perfect pair transmission\,\cite{Golubov99,Lofwander01} and should lead to a finite shot noise\,\cite{Lofwander02,Lofwander03}. In the case of tunneling from a point-like STM tip, in which ZBCP are often observed\,\cite{sharoni02,sharoni03}, the geometry is very far from planar, so it is hard to expect $\Gamma_p$ to be close to unity. Indeed, the differential conductivity of the zero-bias peak in STM experiments\,\cite{sharoni02,sharoni03,kato1,kato2,Davis01} is typically well below $4e^2/h$ expected for a single channel with unit transmission.

Below, we report conductance and noise measurements in NS junctions made on (110) plane of optimally doped (x=15\%) \LSCO{} (LSCO). The $\Delta(\bf k)$ in LSCO, as in other cuprate HTSC materials, is of $d_{x^2-y^2}$ symmetry\,\footnote{we use here the standard axes designation with x, y and z being in the a, b, and c directions respectively}.  Therefore, for electrons incident on (110) plane, $\Delta(-k_\perp)=-\Delta(k_\perp)$ and the zero energy states (ZES) are formed for every direction of an incident electron. As a result, the differential conductance should exhibit ZBCP. Transport through the ZES can be either due to Andreev reflection with charge $2e$ transferred in a single event, or due to a single-particle process, in which an electron coming from the normal metal picks up a pair from the HTSC to enter the ZES\,\cite{Lofwander02,Lofwander03}. We exploit shot noise sensitivity to the transmitted charge to distinguish between these processes.

\begin{figure}
\includegraphics[keepaspectratio=true,scale=0.30]{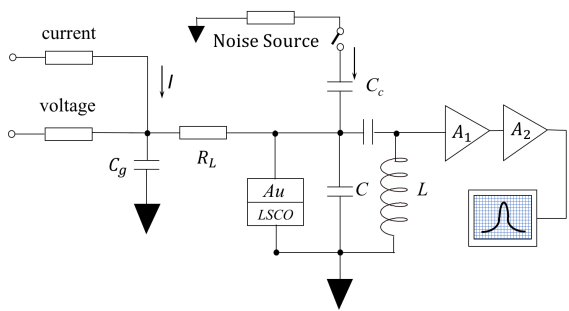}
\caption{The measurement circuit. $A_1$ and $A_2$ are the cryogenic and the room-temperature amplifiers respectively. DC current is driven through the load resistor $R_L$, left hand side of which is effectively grounded at RF frequencies through capacitor $C_g$. Noise source was used for calibration at each value of the temperature and the current through a junction.}
\label{Fig:circuit}
\end{figure}

We used experimental setup shown schematically in Fig.\,\ref{Fig:circuit}. Current fluctuations generated by the junction produce voltage fluctuations across the parallel RLC resonant circuit\,\cite{dePicciotto98} at the front end of the cryogenic amplifier. The frequency $f$ of the resonance circuit is determined by the inductance $L$ of the coil, and the capacitance $C$ of the junction and the cables. We chose $f\approx 20$\,MHz to compromise between $1/f$ noise of the junction and current noise of the amplifier, which scales as $f^2$. The width of the resonance $\Delta f=1/2\pi R_{\parallel}C$, where $R_{\parallel}^{-1}=R_{ac}^{-1}+R_L^{-1}$ depends on the load $R_L$ and junction $R_{ac}$ resistances at the resonance frequency.  The signal from the cryogenic amplifier is fed into a room temperature one followed by the spectrum analyzer with resolution bandwidth either 10\,KHz or 100\,KHz, depending on $\Delta f$. The resonant circuit, supplementary resistors and capacitors, and the cryogenic amplifier were mounted on the 2.7\,K stage of a cryo-free system inside the vacuum chamber.  The sample box was anchored to the stage with a heat sink, weak enough to allow the sample heating without significantly affecting the stage temperature; the sample box temperature was measured with a calibrated diode. The components of the measurement circuit were mounted inside a copper cage to minimize pickup noise. For noise measurements and calibration we used semirigid coaxial cables.
\noindent
\vskip 0.05in

We prepared the NS junctions by sputtering gold on mechanically, and then chemically polished surface of a LSCO single crystal. Such a surface is covered by a natural insulating layer, which serves as a barrier. We defined ${\rm 50\times 50\mu m^2}$ gold pads by lithography and contacted them with wire bonds. Gold, sputtered on the back side of the samples and annealed at 500$^\circ$\,C, served as the ground contact. We measured the low-frequency differential conductance $G_{dc}=dI/dV$  typically at 77\,Hz. The differential conductance at the frequency of the noise measurements, $G_{ac}$ was obtained from the width of the resonance; it happened to be higher than $G_{dc}$ by about 10-15\%, depending on the junction resistance; this discrepancy was important for proper calibration of the noise source,  which was done against the thermal noise of the junction, similarly to Ref.\,\cite{dePicciotto98}.

\begin{figure}[h]
\includegraphics[keepaspectratio=true,scale=0.45]{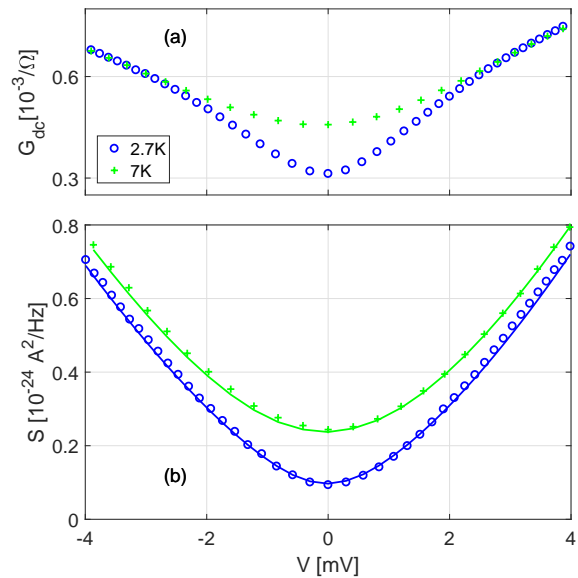}
\caption{(a) Differential conductance of a high resistance  \LSCO{} junction J1. The $G_{dc}$ is V-shaped and the noise in pannel (b) is fitted well with charge $e$, Eq.\,\ref{Eq:coth}. }
\label{Fig:J1}
\end{figure}

In Fig.\,\ref{Fig:J1}(a) we show $G_{dc}$ for a high resistance junction J1 on (110) surface, and in Fig.\,\ref{Fig:J1}(b) the noise generated by it.
The junction, as prepared, did not show zero-bias conductance peak due to low transmission, $\Gamma_e\ll 1$. The differential conductance is V-shaped, and the noise generated by the junction agrees well without any fitting parameter with the prediction\,\cite{RogovinScalapino,LevitovReznikov05}:

\be S=2qI\coth(qV/2k_BT)\label{Eq:coth}\ee
with charge $q=e$\emph{}. Note, that Eq.\,\ref{Eq:coth} is valid even for nonlinear junctions provided the transmission probability through the barrier is small.

\begin{figure}
\includegraphics[keepaspectratio=true,scale=0.45]{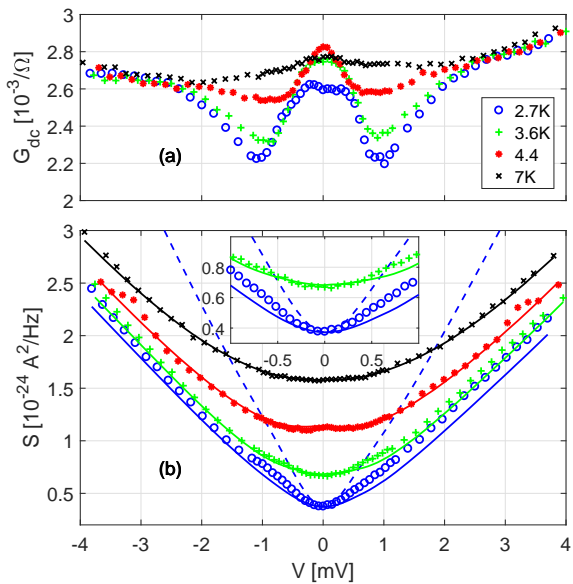}
\caption{(a) Conductance of junction J2 after discharge at 80\,V. The ZBCP and minima on its sides are suppressed with heating at temperatures much smaller than the bulk \tc. (b) experimental results for the noise, and the fit using  Eq.\,\ref{Eq:coth2} with $\G=0.4$ for $V>0$ and $0.3$ for $V<0$;  $S_e$ -- solid lines,  and $S_{2e}$ for $T=2.7$\,K -- dashed line. Inset: low voltage blowup.}
\label{J2}
\vskip-.05in
\end{figure}

In order to improve the transmission we, following\,\cite{dagan99}, created pinholes in the surface barrier by discharging a 220\,nF capacitor through the junction while keeping it cold. The resistance of a junction would decrease with increase in charging voltage $V_c$, until eventually would jump up and could not be further decreased; we considered such a junction as being burned. For junction J2 we started to see ZBCP  at $V_c=70$\,V. In Fig.\,\ref{J2}(a) we show the conductance of the junction after discharge at 80\,V. The ZBCP is slightly split at 2.7\,K, and is suppressed with heating already at $\rm T\approx7$\,K, well below the bulk \tc$\approx 35$\,K.

\vskip .15in
\noindent
\vskip .05in

The noise $S$ generated by the junction is shown in Fig.\,\ref{J2}(b). At low temperature and voltages it lies above the expectation  for the single-electron transport through the junction. Importantly, the noise is not accompanied by the large fluctuations $\delta S$ of the noise power, sometimes seen at larger voltages across the junctions\,\footnote{Sometimes, especially with low-transmission barriers, we see excess noise at voltages 2-4\,mV across the junctions. Such noise is accompanied by the large fluctuations $\delta S$, and will be published elsewhere.}.

Eq.\,\ref{Eq:coth} cannot be used to calculate the expected noise, since the junction after the discharge is no longer a low-transmission one. This is apparent from the noise at $eV\gg k_BT$, which is temperature dependent, and its Fano factor $F=(1/2e)dS/dI\approx 0.6$ is significantly smaller than one. Note, that Eq.\,\ref{Eq:coth} predicts temperature-independent noise with $F=1$ at $eV\gg k_BT$.
Therefore, we choose to compare the observed noise with a generalization of the prediction for non-interacting electrons with constant transmission through the barrier\,\cite{Blanter00}:

\be S=2qI(1-\G)\coth(qV/2k_BT)+2k_BT\G (G(0)+G(V))\label{Eq:coth2}\ee
\\
\noindent where $\G=\langle\Gamma_i^2\rangle/\langle\Gamma_i\rangle$, $\Gamma_i$ is the transmission probability through the barrier for a channel $i$. Formula~\ref{Eq:coth2}, derived in the supplementary material, does not have the generality of\,\ref{Eq:coth}. Still, it (i) reduces to the exact one for noninteracting electrons and energy independent $\Gamma_i$; (ii) gives correct temperature dependence for energy-dependent $\Gamma_i$ at $eV\gg k_BT$; (iii) has correct zero-voltage ($4TG$) and low-transmission (Eq.\,\ref{Eq:coth}) limits. In addition, it correctly captures interaction-induces charge renormalization for $\Gamma_p\ll1$ and energy-independent $\Gamma_p$ cases\,\cite{Rogovin1974,Khlus1987,LevitovReznikov05,Muzy}.

\noindent

The noise $S_e$ expected from Eq.\,\ref{Eq:coth2} with $q=e$ is shown in Fig.\,\ref{J2}(b); we used $\G=0.4$ (corresponding to $F=0.6$) for $V>0$ and $\G=0.3$ ($F=0.7$) for $V<0$ to fit the data at large voltages, where the transmission is, presumably, constant. At temperatures $T=7$\,K   and 4.4\,K the fit is good, it even reproduces a small maximum at $V=0$ due to the ZBCP. At lower temperatures, the experimental data lies above this expectation, with largest deviation at the lowest temperature of 2.7\,K. We plot for comparison the noise $S_{2e}$ expected for $q=2e$ and the same $\G$. The experimental data lies in between the $S_e$ and $S_{2e}$ expectations, indicating that part of the noise is generated by Andreev reflection; the rest is due to the single-particle transport. Most of the pair contribution is accumulated below the crossover voltage $|\tilde V|<0.5$\,mV. At $|V|>\tilde V$ the noise increases with voltage roughly as predicted by Eq.\,\ref{Eq:coth2} with $q=e$.

\noindent
\begin{figure}
\makebox[\linewidth]{
\includegraphics[keepaspectratio=true,scale=0.45]{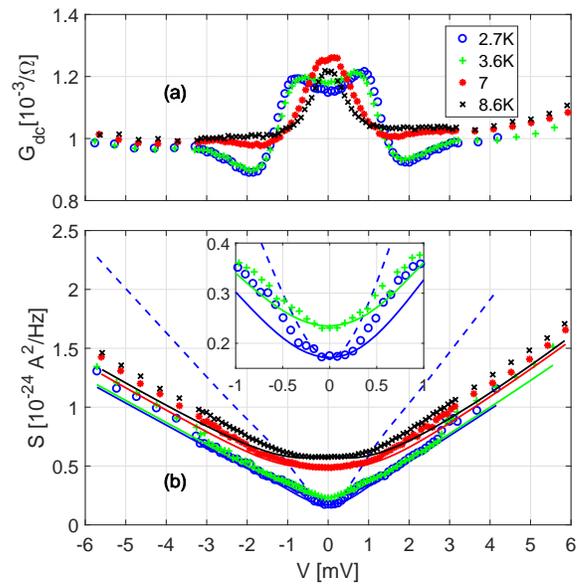}}
\caption{(a) conductance and (b) noise for junction J3 after discharge at 70\,V; notations are the same as in Fig.\,\ref{J2}. In this junction ZBCP is twice as wider as in J2 and pronouncedly split. $\G=0.3$ for $V>0$ and $0.4$ for $V<0$. }
\label{J3}
\end{figure}

An interesting question is what sets the scale of $e\tilde V$, which is much smaller than $k_B$\tc{}. It is conspicuously similar to the ZBCP half-width. However, this seems to be coincidental: we got Cooper-pair generated noise with four optimally doped samples; all of them with similar $\tilde V$ but different ZBCP width.  As an example, we show in Fig.\,\ref{J3} the results for junction J3, in which ZBCP is twice as wide as in J2.

In order to quantify the pair contribution to the noise we calculated $\nu=(S-S_e)/(S_{2e}-S_e)$  for $|V|<|\tilde V|$ which we found for junction J2 to be 26\% at 2.7\,K and 17\% at 3.7\,K; for junction J3 $\nu=51$\% and 31\% respectively; similar values were obtained for other junctions. It is not surprising that only part of the electrons passes the junction through Andreev reflection.  Indeed, it is possible for a single electron to cross the barrier. In a d-wave superconductor electrons moving in the nodal direction can pass into the unpaired quasiparticle states. Alternatively, an electron from the normal metal can be thermally activated above the gap and go into a quasiparticle state; such a process occurs in BCS superconductors at finite temperatures; it leads to conductance increase with temperature for the low-transmission junctions. Finally an electron can pair with a one from  the quasiparticle state and entry the ZES at the surface.

It is natural to expect the pair contribution to the current to be more pronounced in the ZBCP region, since the interference of the waves with opposite sign of $k_\perp$ impinging at (110) surface increases Andreev reflection probability. However, the pair contribution disappears at temperatures well below the temperature of the ZBCP disappearance, see Figs.\,\ref{J2},\ref{J3}. The similarity of $e\tilde V/k_B\approx 6$\,K and the temperature $T\approx 4$\,K, at which the pair-generated noise disappears within our experimental accuracy, sets a new energy scale. Such a scale can be attributed to an  additional order parameter component with $i{d_{xy}}$ or $is$ symmetry. Such an addition can exist at a surface, especially at (110) surface, since it breaks inversion symmetry, and therefore suppresses $d_{x^2-y^2}$ order parameter. Ginzburg-Landau type arguments show that this suppression can lead to an appearance of an additional order parameter, which is less energetically favorable than $\Delta_{x^2-y^2}$ in the bulk\,\cite{Laughlin95} but wins at the surface. This additional small order parameter is nonzero in any direction, and therefore can block single-particle transport through the junction, leading to an increased relative contribution of the Cooper pairs to the current. Since it is destroyed at very low temperatures, such an additional order parameter would mean time-reversal symmetry breaking very different from the one which may exist in pseudogap regime\,\cite{Kapitulnik08}. The current through the junction is concentrated at the pinholes created by the discharge therefore, the required symmetry breaking can be local\,\cite{Bourges06} and would not be detectable in experiments like Kerr rotation\,\cite{Kapitulnik08}, which average over a large area.

In agreement with expectations, we observe no ZBCP in junctions on (100) surface of the optimally doped \LSCO{}. The observed shot noise agrees well with the one generated by single-electron transport, since, in the absence of transmission enhancement leading to ZBCP, it totally dominates the conductance through the junctions.  Presented results, to the best of our knowledge, are the first observation of Cooper pairs in cuprate superconductors via noise measurement.

We thank D.\,Podolsky and A.\,Auerbach for discussion. This work was supported by Israeli Science Foundation.

\section{Supplementary material}
Below we shall derive Eq.\,\ref{Eq:coth2} of the paper for the thermal and shot noise generated by electrons partitioned by a barrier\,\footnote{ We define the spectral density for the positive frequencies only.}. Let us consider a multi-channel junction separated by a barrier with energy and channel-dependent transmissions $\Gamma_n(E)$. In the case there is no correlation between different channels, the current trough the junction is given by:

\begin{equation} \label{S1}
 I = \frac{e}{h} \sum_n \int dE   \Gamma_n(E)\{\left( f_L (1 - f_R) - f_R (1 - f_L) \right)
\end{equation}
Here $f_L=(1+e^{(E-eV)/K_B T)})^{-1}$ and $f_R=(1+e^{E/K_B T)})^{-1}$ are the Fermi distribution functions on the left and right sides of the barrier, and $V$ is the voltage across the barrier.

In order to calculate low-frequency spectral density of the noise, we shall start with Eq.\,61 of Ref.\,\cite{Blanter00},  which can be written as a sum of two terms $S=S_1+S_2$:

\begin{eqnarray} \label{S1}
\begin{aligned}
S_1 = &\frac{2e^2}{h} \sum_n \int dE   \Gamma_n(E)\left(1-\Gamma_n(E)\right)\times\\
&\times\left( f_L (1 - f_R) + f_R (1 - f_L) \right)
\end{aligned}
\end{eqnarray}

\begin{eqnarray} \label{S2}
\begin{aligned}
S_2=&\frac{2e^2}{h} \sum_n \int dE\Gamma_n(E)^2\times \\
&\times\left( f_L(1-f_L) + f_R(1-f_R) \right)
\end{aligned}
\end{eqnarray}

\noindent At $eV\gg K_B T$ term $S_1$ is almost temperature independent and can be loosely viewed the shot noise, and term $S_2$ is proportional to the temperature and constitutes the thermal contribution to the noise; at low voltages, there is no such a distinction.

We observe that $f_R (1 - f_L)=e^{-eV/K_B T} f_L (1 - f_R)$, and therefore the expressions for $I$ and $S_1$ would be very similar, if not for the factor $1-\Gamma_n(E)$ in $S_1$. In order to proceed, we define

\be\label{G} \G=\sum_n \Gamma_n^2/\sum_n\Gamma\ee

\noindent and assume $\G$ to be energy independent. Now the integral in Eq.\,\ref{S1} can be written as:

\begin{eqnarray}\label{S1.2}
 \nonumber S_1 &= \frac{2e^2}{h} (1-\G) \int dE   \Gamma_n(E)  f_L (1 - f_R)(1+e^{-eV/K_B T})=\\
 &=2eI(1-\G)\coth(eV/2K_B T)
\end{eqnarray}

For a low transmission junction, Eq.\,\ref{S1.2} reduces to Eq.\,1 of the paper for the charge $q=e$. The main contribution to $S_2$ comes from the narrow regions of width $K_B T$ around $E=0$ and $E=eV$, and therefore can be approximated as

\begin{eqnarray}\label{S2.2}
\begin{aligned}
S_2=&2K_B T\G \sum_n\frac{e^2}{h}(\Gamma_n(0)+\Gamma_n(eV))=\\
=&2K_B T\G(G(0)+G(V))
\end{aligned}
\end{eqnarray}

\noindent Here $G(V)$ is the voltage-dependent differential conductance of the junction. By this we recover Eq.\,2 of the paper for $q=e$.

There is no contradiction between voltage-dependent conductance $G(V)$ and a constant $\G$. Such a possibility can arise e.g.  for a barrier with energy-independent transmission but energy-dependent density of states on one side of it; this would lead to the energy-dependent number of channels in Eq.\,\ref{G}. This is a plausible situation for our junctions, in which the quasiparticle density of states in the nodal direction increases linearly with voltage.
We note that $\G=2/3$ was predicted for a diffusive conductor on the basis of the random matrix theory\,\cite{dorokhov82}; this fact was used in\,\cite{BeenButt92} to calculate the shot noise. $\G$ between 0 and 0.4, similar to what we used to fit the data, was observed in\,\cite{Tikhonov2013} for transport through a barrier in GaAs structures.

\bibliography{2e}
\end{document}